\documentclass[english,11pt,draftcls,onecolumn]{IEEEtran}
\usepackage[T1]{fontenc}
\usepackage[latin9]{inputenc}
\usepackage{bm}
\usepackage{amsthm}
\usepackage{amsmath}
\usepackage{amssymb}
\usepackage{graphicx}
\usepackage{esint}

\makeatletter
\theoremstyle{plain}
\newtheorem{thm}{\protect\theoremname}
\theoremstyle{plain}
\newtheorem{prop}{\protect\propositionname}

\makeatother

\usepackage{babel}
\providecommand{\propositionname}{Proposition}
\providecommand{\theoremname}{Theorem}

\begin{document}

\title{Optimality of Received Energy in Decision Fusion over Rayleigh Fading
Diversity MAC with Non-Identical Sensors}

\author{Domenico~Ciuonzo,~\IEEEmembership{Student~Member,~IEEE,} Gianmarco~Romano,~\IEEEmembership{Member,~IEEE,}
\\
Pierluigi~Salvo~Rossi,~\IEEEmembership{Senior Member,~IEEE} %
\thanks{The authors are with the Department of Information Engineering, Second
University of Naples, via Roma, 29, 81031 Aversa (CE), Italy. Email:
\texttt{\{domenico.ciuonzo, gianmarco.romano, pierluigi.salvorossi\}@unina2.it}
.%
}\vspace{-2cm}
}
\maketitle
\begin{abstract}
Received-energy test for non-coherent decision fusion over a Rayleigh
fading multiple access channel (MAC) without diversity was recently
shown to be optimum in the case of conditionally mutually independent
and identically distributed (i.i.d.) sensor decisions under specific
conditions \cite{Berger2009,Li2011}. Here, we provide a twofold generalization,
allowing sensors to be non identical on one hand and introducing diversity
on the other hand. Along with the derivation, we provide also a general
tool to verify optimality of the the received energy test in scenarios
with correlated sensor decisions. Finally, we derive an analytical
expression of the effect of the diversity on the large-system performances,
under both individual and total power constraints.\vspace{-0.5cm}

\end{abstract}

\section{Introduction}

Starting from classical distributed detection \cite{Varshney1996},
large efforts in the recent literature have been devoted to the implementation
of distributed detection in wireless sensor networks (WSNs) \cite{Chen2006,Gastpar2006,Chamberland2007}.
Local decisions in a WSN are usually transmitted to a decision fusion
center (DFC) in order to improve reliability of geographically distributed
sensing through central processing. Common system architectures make
reference to the availability of parallel (non-interfering) channels
from the sensors to the DFC \cite{Chen2004,Niu2006a,Lei2010}. However,
more sophisticated setups have been investigated, where the intrinsically
interfering nature of the wireless channel is exploited and not combated
\cite{Berger2009,Zhang2008,Ciuonzo2012}.

Recently, in \cite{Berger2009} and \cite{Li2011}, the received-energy
test was studied for non-coherent decision fusion over a multiple
access channel (MAC). More specifically, in \cite{Berger2009} the
received energy was claimed as optimal for the no-diversity case with
conditionally (given the phenomenon) mutually independent and identically
distributed (i.i.d.) sensor local decisions, as long as the probability
of false alarm of the generic sensor is lower than the corresponding
probability of detection. Also, analytical performances of the received-energy
test in the diversity scenario were derived. However, optimality property
of the test was not investigated. The optimality of the test for the
no-diversity case with conditionally i.i.d. sensor local decisions
was proven in \cite{Li2011}. Only the case with sensors whose probability
of false alarm is lower than the corresponding probability of detection
was considered. Nonetheless, the diversity case was still ignored
in the optimality analysis.

The main contributions of this correspondence are:
\begin{itemize}
\item a rigorous proof of the \emph{optimality} of the received-energy test
for non-coherent decision fusion%
\footnote{Although, energy receiver and non-coherent are not synonyms, in the
paper we will confuse them. In the related literature, such a misuse
is common due to the fact that the energy detector is the default
receiver adopted for non-coherent decision fusion.%
} over a Rayleigh fading MAC with \emph{arbitrary order of diversity}
and with conditionally mutually independent \emph{but non identically
distributed} (i.n.i.d.)\emph{ }sensor decisions\emph{, }as long as
\emph{each} sensor probability of false alarm is lower than the correspondent
probability of detection\emph{;}
\item as a side result, a sufficient condition on the log-likelihood ratio
(LLR) of the number of active sensors suited for testing received
energy optimality in scenarios with correlated local decisions;
\item analytical derivation of large-system performances for conditionally
i.i.d. sensor local decisions as a function of the order of diversity,\emph{
}where two different scenarios are considered: (a) sensors with an
individual power constraint (IPC); (b) sensors with a total power
constraint (TPC).
\end{itemize}
It is worth noting that in \cite{Ciuonzo2012} a different scenario
was analyzed, where: ($i$) conditionally i.i.d. sensor decisions
were considered, and ($ii$) instantaneous channel state information
(CSI) at the DFC was assumed. The focus was on the performance analysis
of several sub-optimal fusion rules in terms of complexity, required
knowledge, probability of detection and false alarm.

The paper is organized as follows: Sec. \ref{sec:System-Model} introduces
the system model; in Sec. \ref{sec:Optimality-RE} we present the
main results of this correspondence, while in Sec. \ref{sec:Concluding-Remarks}
we draw some concluding remarks; proofs are confined to Appendices.

\emph{Notation} - Lower-case (resp. Upper-case) bold letters denote
vectors (resp. matrices), with $a_{n}$ (resp. $a_{n,m}$) representing
the $n$th (resp. the $(n,m)$th) element of the vector $\bm{a}$
(resp. matrix $\bm{A}$); upper-case calligraphic letters, e.g. $\mathcal{A}$,
denote discrete and finite sets; $\bm{I}_{N}$ denotes the $N\times N$
identity matrix; $\bm{0}_{N}$ (resp. $\bm{1}_{N}$) denotes the null
(resp. ones) vector of length $N$; $\mathbb{E}\{\cdot\}$, $(\cdot)^{t}$,
$(\cdot)^{\dagger}$, $\Re\left(\cdot\right)$, $\Im(\cdot)$ and
$\left\Vert \cdot\right\Vert $ denote expectation, transpose, conjugate
transpose, real part, imaginary part and Frobenius norm operators;
$P(\cdot)$ and $p(\cdot)$ are used to denote probability mass functions
(pmf) and probability density functions (pdf), while $P(\cdot|\cdot)$
and $p(\cdot|\cdot)$ their corresponding conditional counterparts;
$\mathcal{N}_{\mathbb{C}}(\bm{\mu},\bm{\Sigma})$ (resp. $\mathcal{N}(\bm{\mu},\bm{\Sigma})$)
denotes a circular symmetric complex (resp. real) normal distribution
with mean vector $\bm{\mu}$ and covariance matrix $\bm{\Sigma}$,
$\mathcal{B}(k,p)$ denotes a binomial distribution of $k$ trials
with probability of success $p$ and $\chi_{L}^{2}$ denotes a chi-square
distribution with $L$ degrees of freedom; $(a*b)(\ell)$ denotes
the convolution between series $a(\ell)$ and $b(\ell)$; finally
the symbols $\sim$ and $\overset{d}{\rightarrow}$ mean \textquotedblleft{}distributed
as\textquotedblright{} and ``convergence in distribution''.

\section{System Model\label{sec:System-Model}}

\subsection{WSN modeling}

We consider a distributed binary hypothesis test, where $K$ sensors
are used to discriminate between the hypotheses of the set $\mathcal{H}=\{H_{0},H_{1}\}$,
representing, not necessarily, the absence ($H_{0}$) or the presence
($H_{1}$) of a specific target of interest. The $k$th sensor, $k\in\mathcal{K}\triangleq\{1,2,\ldots,K\}$,
takes a binary local decision $d_{k}\in\mathcal{H}$ about the observed
phenomenon on the basis of its own measurements. 

Each decision $d_{k}$ is mapped to a symbol $x_{k}\in{\cal X}=\{0,1\}$
representing an On-Off Keying (OOK) modulation: without loss of generality
we assume that $d_{k}=H_{i}$ maps into $x_{k}=i$, $i\in\{0,1\}$.
The quality of the $k$th sensor decisions is characterized by the
conditional probabilities $P(x_{k}|H_{j})$. More specifically, we
denote $P_{D,k}\triangleq P\left(x_{k}=1|H_{1}\right)$ and $P_{F,k}\triangleq P\left(x_{k}=1|H_{0}\right)$,
respectively the probability of detection and false alarm of the $k$th
sensor.

The sensors communicate with the DFC over a wireless flat-fading MAC,
modeled through i.i.d. Rayleigh fading coefficients with equal mean
power. The DFC employs an $N$-diversity approach in order to combat
signal attenuation due to small-scale fading of the wireless medium.
The diversity can be accomplished with time, frequency, code or antenna
diversity (as recently proposed in \cite{Zhang2008,Banavar2010}).
Statistical CSI is assumed at the DFC, i.e. only the pdf of each fading
coefficient is available.

We denote: $y_{n}$ the received signal at the $n$th diversity branch
of the DFC after matched filtering and sampling; $h_{n,k}\sim{\cal N}_{\mathbb{C}}\left(0,\sigma_{h}^{2}\right)$
the fading coefficient between the $k$th sensor and the $n$th diversity
branch of the DFC%
\footnote{It is worth noting that assuming an asymmetric model for channel coefficient
statistics would be more realistic. However, this would make the results
much more dependent on the specific scenario without adding any significant
insight from a theoretical point of view. A symmetric model for channel
coefficient statistics is assumed for a two-fold reason: on one side
it can be considered as a starting point before analyzing more realistic
application-dependent scenarios; on the other side a symmetric scenario
could represent scenarios in which power control is considered.%
}; $w_{n}$ the additive white Gaussian noise at the $n$th diversity
branch of the DFC. The vector model at the DFC is the following:
\begin{equation}
\bm{y}=\bm{H}\bm{x}+\bm{w}\label{eq:channel_model}
\end{equation}
where $\bm{y}\in\mathbb{C}^{N}$, $\bm{H}\in\mathbb{C}^{N\times K}$,
$\bm{x}\in\mathcal{X}^{K}$, $\bm{w}\sim\mathcal{N}_{\mathbb{C}}(\bm{0}_{N},\sigma_{w}^{2}\bm{I}_{N})$
are the received signal vector, the channel matrix, the transmitted
signal vector and the noise vector, respectively. Finally, we define
the random variable $\ell\triangleq\ell(\bm{x})=\sum_{k=1}^{K}x_{k}$,
representing the number of active sensors and the set $\mathcal{L}\triangleq\{0,\ldots,K\}$
of possible realizations of $\ell$.

\subsection{LLR}

The optimal test \cite{Varshney1996,Kay1998} for the considered problem
can be formulated as 
\begin{equation}
\Lambda_{opt}\triangleq\ln\left[\frac{p(\bm{y}|H_{1})}{p(\bm{y}|H_{0})}\right]\begin{array}{c}
{\scriptstyle \hat{H}=H_{1}}\\
\gtrless\\
{\scriptstyle \hat{H}=H}_{0}
\end{array}\gamma\label{eq:neyman_pearson_test}
\end{equation}
where $\hat{H}$, $\Lambda_{opt}$ and $\gamma$ denote the estimated
hypothesis, the Log-Likelihood-Ratio (LLR, i.e. the optimal fusion
rule) and the threshold to which the LLR is compared to. The threshold
$\gamma$ can be determined to assure a fixed system false-alarm rate
(Neyman-Pearson approach) or can be chosen to minimize the probability
of error (Bayes approach) \cite{Varshney1996,Kay1998}. An explicit
expression of the LLR from Eq.~(\ref{eq:neyman_pearson_test}) is
given by
\begin{eqnarray}
\Lambda_{opt} & = & \ln\left[\frac{\sum_{\ell=0}^{K}p(\bm{y}|\ell)P(\ell|H_{1})}{\sum_{\ell=0}^{K}p(\bm{y}|\ell)P(\ell|H_{0})}\right]=\ln\left[\frac{\sum_{\ell=0}^{K}\frac{1}{(\sigma_{w}^{2}+\ell\sigma_{h}^{2})^{N}}\exp\left(-\frac{\bm{\|y}\|^{2}}{\sigma_{w}^{2}+\ell\sigma_{h}^{2}}\right)P(\ell|H_{1})}{\sum_{\ell=0}^{K}\frac{1}{(\sigma_{w}^{2}+\ell\sigma_{h}^{2})^{N}}\exp\left(-\frac{\bm{\|y}\|^{2}}{\sigma_{w}^{2}+\ell\sigma_{h}^{2}}\right)P(\ell|H_{0})}\right]\label{eq:optimum_llr}
\end{eqnarray}
where we have exploited the conditional independence of $\bm{y}$
from $H_{i}$ (given $\ell$). 

In the case of conditionally (given $H_{i}$) i.i.d. sensor decisions
($(P_{D,k},P_{F,k})=(P_{D},P_{F})$, $k\in\mathcal{K}$) we have that
$\ell|H_{1}\sim\mathcal{B}(K,P_{D})$ and $\ell|H_{0}\sim\mathcal{B}(K,P_{F})$.
Differently, when local sensor decisions are conditionally i.n.i.d.
the pmfs $P(\ell|H_{i})$ are represented by the more general \emph{Poisson-Binomial}
distribution \cite{Belfore1995,Chen1997,Fernandez2010}, with expressions
given by:
\begin{eqnarray}
P(\ell|H_{1}) & = & \sum_{\bm{x}:\ell(\bm{x})=\ell}\prod_{k=1}^{K}(P_{D,k})^{x_{k}}\prod_{s=1}^{K}(1-P_{D,s})^{(1-x_{s})}\label{eq:Poisson_Binomial_explicit}\\
P(\ell|H_{0}) & = & \sum_{\bm{x}:\ell(\bm{x})=\ell}\prod_{k=1}^{K}(P_{F,k})^{x_{k}}\prod_{s=1}^{K}(1-P_{F,s})^{(1-x_{s})}
\end{eqnarray}
It is worth noting that Eq. (\ref{eq:Poisson_Binomial_explicit})
requires sums which are infeasible to compute in practice unless the
number of sensors $K$ is small. For this reason different methods
have been proposed in literature for its efficient evaluation. The
alternatives include fast convolution of individual Bernoulli pmfs
\cite{Belfore1995}, recursive approaches \cite{Chen1997} and a Discrete
Fourier Transform (DFT) based computation \cite{Fernandez2010}.

\section{Optimality of Received Energy\label{sec:Optimality-RE}}

As already stated in \cite{Berger2009}, the received energy $\psi\triangleq\left\Vert \bm{y}\right\Vert ^{2}$
is a sufficient statistic for the LLR, since Eq.~(\ref{eq:optimum_llr})
depends on $\bm{y}$ only through $\psi$. However, \emph{sufficiency
}alone does not guarantee that the test
\begin{equation}
\psi\begin{array}{c}
{\scriptstyle \hat{H}=H_{1}}\\
\gtrless\\
{\scriptstyle \hat{H}=H}_{0}
\end{array}\gamma'\label{eq:Energy Detection}
\end{equation}
is equivalent to Eq.~(\ref{eq:neyman_pearson_test}). As shown in
\cite{Li2011}, the test in Eq.~(\ref{eq:Energy Detection}) is optimal
\emph{iff} $\Lambda_{opt}(\psi)$ is a strictly increasing function
of $\psi$. If this property is satisfied, the test in Eq.~(\ref{eq:neyman_pearson_test})
is equivalent to Eq.~(\ref{eq:Energy Detection}) by simply setting
$\gamma'=\Lambda_{opt}^{-1}(\gamma)$. For this purpose in the following
we first introduce a general optimality test (in the form of a sufficient
condition) which relates the pmfs $P(\ell|H_{i})$, $H_{i}\in\mathcal{H}$,
to assure that $\Lambda_{opt}(\psi)$ is strictly increasing in the
case of an $N$-diversity MAC.
\begin{prop}
A sufficient condition for $\Lambda_{opt}(\psi)$ to be a strictly
increasing function of $\psi$ is given by:\label{prop:suff_condition_for_optimality}
\begin{equation}
\lambda(\ell)>\lambda(\ell-1),\quad\ell\in\mathcal{L}\backslash\{0\}\label{eq:RFS-LLRmonotonicity}
\end{equation}
where $\lambda(\ell)\triangleq\ln\left[\frac{P(\ell|H_{1})}{P(\ell|H_{0})}\right]$.\end{prop}
\begin{IEEEproof}
The proof is reported in Appendix \ref{sec:Appendix_Proposition-Suffcond}.
\end{IEEEproof}
The above proposition states that strictly increasing property of
$\lambda(\ell)$ assures optimality of the test in Eq.~(\ref{eq:Energy Detection}).
We will refer to $\lambda(\ell)$ as the $\ell-$LLR hereinafter%
\footnote{Note that we will not consider in Eq.~(\ref{eq:RFS-LLRmonotonicity})
(and throughout the paper) the case $\ell=0$ when testing $\ell$-LLR
strictly increasing property, since $\lambda(-1)$ has no physical
meaning.%
}. It is worth noting that such a sufficient condition is independent
of the order of diversity $N$. Also, Eq.~(\ref{eq:RFS-LLRmonotonicity})
depends on the WSN model only through the number of active sensors
$\ell$ (given $H_{i}$) and does not require any specific assumption
on $P(\bm{x}|H_{i})$, e.g. conditional mutual independence of local
sensor decisions, i.e. $P(\bm{x}|H_{i})=\prod_{k=1}^{K}P(x_{k}|H_{i})$.
This means that Eq.~(\ref{eq:RFS-LLRmonotonicity}) plays the role
of a general property for received energy optimality, to be verified
even in the case of conditionally correlated local sensor decisions.

In the simplest case of conditionally i.i.d. local sensor decisions,
($P_{D,k},P_{F,k})=(P_{D},P_{F})$, $k\in\mathcal{K}$, as assumed
in \cite{Berger2009,Li2011}, the strictly increasing property of
$\lambda(\ell)$, $\ell\in\mathcal{L}$, is equivalent to 
\begin{eqnarray}
\frac{\tbinom{K}{\ell}P_{D}^{\ell}(1-P_{D})^{K-\ell}}{\tbinom{K}{\ell}P_{F}^{\ell}(1-P_{F})^{K-\ell}} & > & \frac{\tbinom{K}{\ell-1}P_{D}^{\ell-1}(1-P_{D})^{K-\ell+1}}{\tbinom{K}{\ell-1}P_{F}^{\ell-1}(1-P_{F})^{K-\ell+1}},\label{eq:Lllr_increasing_ciid}
\end{eqnarray}
that reduces to $P_{D}>P_{F}.$ This result, not only confirms theoretical
findings for optimality of Eq.~(\ref{eq:Energy Detection}) when
$N=1$ as in \cite{Berger2009,Li2011}, but it also proves the optimality
of the test over the \emph{Diversity MAC }(i.e. $N>1$)\emph{ }used
in \cite{Berger2009}; this result shows the effectiveness and simplicity
of Proposition \ref{prop:suff_condition_for_optimality} w.r.t. the
approach taken in \cite{Li2011}.

Differently, when sensor decisions are conditionally i.n.i.d. (i.e.
the case of a heterogeneous WSN), the following theorem generalizes
the result in Eq.~(\ref{eq:Lllr_increasing_ciid}).
\begin{thm}
If $P(\bm{x}|H_{i})=\prod_{k=1}^{K}P(x_{k}|H_{i})$, $H_{i}\in\mathcal{H}$,
and $P_{D,k}>P_{F,k}$, $k\in\mathcal{K}$, the $\ell$-LLR satisfies
the strictly increasing property described in Eq.~(\ref{eq:RFS-LLRmonotonicity})
and thus Eq.~(\ref{eq:Energy Detection}) is the optimal test when
an $N$-diversity MAC is employed.\label{thm:RFS_monotonicity}\end{thm}
\begin{IEEEproof}
The proof is reported in Appendix \ref{sec:Proposition_Monotonic LLR}.
\end{IEEEproof}
Theorem \ref{thm:RFS_monotonicity} states that under non identical
sensors $(P_{D,k},P_{F,k})$ and $N$-diversity MAC the received energy
$\psi$ is again the optimal test. Also, Theorem \ref{thm:RFS_monotonicity}
relies on a sufficient condition, i.e. specific WSN configurations
not satisfying the assumption $P_{D,k}>P_{F,k}$, $k\in\mathcal{K}$,
but still verifying Eq.~(\ref{eq:RFS-LLRmonotonicity}) may exist.
Although such a case is not ``typically'' of interest, since the
condition $P_{D,k}\leq P_{F,k}$ for $k$th sensor is not realistic
in practical scenarios (i.e. sensors operating under nominal conditions),
it proves the robustness of the received energy in scenarios with
some faulty (or byzantine) sensors%
\footnote{A WSN with $K=3$ sensors such that $(P_{D,1},P_{F,1})=(0.5,0.05)$,
$(P_{D,2},P_{F,2})=$(0.4,0.1) and $(P_{D,3},P_{F,3})=(0.3,0.4)$
verifies the property in Eq.~(\ref{eq:RFS-LLRmonotonicity}). %
}.

We finally evaluate analytically the performances as the number of
sensors goes large, in the case of conditionally i.i.d. sensors. Both
IPC and TPC on the WSN and arbitrary diversity $N$ are considered
here. This result generalizes \cite{Li2011}, where no-diversity ($N=1$)
and IPC assumptions were made in deriving formulas. We define 
\begin{equation}
\bm{z}\triangleq\frac{1}{\sqrt{P_{F}K\sigma_{h}^{2}}}\left(\frac{\bm{Hx}}{\sqrt{N}}+\bm{w}\right)\label{eq:model_IPC}
\end{equation}
where, compared to Eq.~(\ref{eq:channel_model}), $\frac{1}{\sqrt{P_{F}K\sigma_{h}^{2}}}$
is a merely scaling factor and $\bm{H}\bm{x}$ is replaced with $\frac{\bm{H}\bm{x}}{\sqrt{N}}$
in $\bm{z}$ in order to keep a fixed amount of average energy $\varepsilon\triangleq E\{\left\Vert \bm{H}\bm{x}\right\Vert ^{2}\}$
w.r.t. $N$. Then we define the system probabilities of false alarm
and detection, respectively $P_{F_{0}}$ and $P_{D_{0}}$, as:
\begin{equation}
P_{F_{0}}\triangleq P(\left\Vert \bm{z}\right\Vert ^{2}\geq\bar{\gamma}|H_{0}),\quad P_{D_{0}}\triangleq P(\left\Vert \bm{z}\right\Vert ^{2}\geq\bar{\gamma}|H_{1}),\label{eq:PD0_PF0_IPC}
\end{equation}

Eqs.~(\ref{eq:model_IPC}) and (\ref{eq:PD0_PF0_IPC}) hold for TPC
scenario when replacing $\bm{z}$ with $\tilde{\bm{z}}\triangleq\frac{1}{\sqrt{P_{F}\sigma_{h}^{2}}}\left(\frac{\bm{Hx}}{\sqrt{KN}}+\bm{w}\right)$.
In this case the average energy is kept fixed w.r.t. both $K$ and
$N$.
\begin{thm}
If $\sigma_{h}^{2}$ and $\sigma_{w}^{2}$ are finite, as $K\rightarrow+\infty$
\begin{eqnarray}
\bm{z}|H_{0}\overset{d}{\rightarrow}\mathcal{N}_{\mathbb{C}}\left(\bm{0}_{N},\frac{1}{N}\bm{I}_{N}\right), & \qquad & \bm{z}|H_{1}\overset{d}{\rightarrow}\mathcal{N}_{\mathbb{C}}\left(\bm{0}_{N},\frac{P_{D}}{P_{F}N}\bm{I}_{N}\right),\\
\tilde{\bm{z}}|H_{0}\overset{d}{\rightarrow}\mathcal{N}_{\mathbb{C}}\left(\bm{0}_{N},\frac{1}{\alpha_{F}}\frac{1}{N}\bm{I}_{N}\right), & \qquad & \tilde{\bm{z}}|H_{1}\overset{d}{\rightarrow}\mathcal{N}_{\mathbb{C}}\left(\bm{0}_{N},\frac{1}{\alpha_{D}}\frac{P_{D}}{P_{F}N}\bm{I}_{N}\right),
\end{eqnarray}
where $\alpha_{F}\triangleq\frac{P_{F}\sigma_{h}^{2}}{P_{F}\sigma_{h}^{2}+\sigma_{w}^{2}N}$
and $\alpha_{D}\triangleq\frac{P_{D}\sigma_{h}^{2}}{P_{D}\sigma_{h}^{2}+\sigma_{w}^{2}N}$.
Then the large-system $(P_{D_{0}-IPC,}^{*}P_{F_{0}-IPC}^{*})$ are
given by:\label{thm:Theorem4}
\begin{eqnarray}
P_{F_{0}-IPC}^{*}(\bar{\gamma}) & = & \exp\left(-\bar{\gamma}N\right)\times\sum_{n=0}^{N-1}\frac{1}{n!}\left(\bar{\gamma}N\right)^{n};\label{eq:asymptotic_perf_PF0}\\
P_{D_{0}-IPC}^{*}(\bar{\gamma}) & = & \exp\left(-\frac{\bar{\gamma}N}{\frac{P_{D}}{P_{F}}}\right)\times\sum_{n=0}^{N-1}\frac{1}{n!}\left(\frac{\bar{\gamma}N}{\frac{P_{D}}{P_{F}}}\right)^{n};\label{eq:asymptotic_perf_PD0}
\end{eqnarray}
while $(P_{D_{0}-TPC,}^{*}P_{F_{0}-TPC}^{*})$ are given by: 
\begin{eqnarray}
P_{F_{0}-TPC}^{*}(\bar{\gamma}) & = & \exp\left(-\bar{\gamma}N\alpha_{F}\right)\times\sum_{n=0}^{N-1}\frac{1}{n!}\left(\bar{\gamma}N\alpha_{F}\right)^{n};\label{eq:asymptotic_perf_PF0_TPC}\\
P_{D_{0}-TPC}^{*}(\bar{\gamma}) & = & \exp\left(-\frac{\bar{\gamma}N\alpha_{D}}{\frac{P_{D}}{P_{F}}}\right)\times\sum_{n=0}^{N-1}\frac{1}{n!}\left(\frac{\bar{\gamma}N\alpha_{D}}{\frac{P_{D}}{P_{F}}}\right)^{n}.\label{eq:asymptotic_perf_PD0_TPC}
\end{eqnarray}
\end{thm}
\begin{IEEEproof}
The proof is reported in Appendix \ref{sec:Proof-of-Theorem 4} for
the IPC case; performance in the TPC scenario can be derived in a
similar fashion.
\end{IEEEproof}
As expected, if $N=1$ the result of Eqs.~(\ref{eq:asymptotic_perf_PF0})
and (\ref{eq:asymptotic_perf_PD0}) coincides with the one given in
\cite[Sec. IV]{Li2011}. 

It is worth remarking that, in both IPC and TPC scenarios with diversity,
Neyman-Pearson and Bayesian error exponents are zero (cfr. with \cite{Li2011}),
because the large-system ROC can not be driven toward the point $(P_{D_{0}}^{*},P_{F_{0}}^{*})=(1,0)$
by increasing the number of sensors, as long as the diversity $N$
is kept finite. This intuition is confirmed by the non-zero values
assumed under an IPC and a TPC by the large-system J-Divergence, $J(p(\bm{z}|H_{0}),p(\bm{z}|H_{1}))=N\times\left[\left(\frac{P_{D}}{P_{F}}+\frac{P_{F}}{P_{D}}\right)-2\right]$,
$J(p(\tilde{\bm{z}}|H_{0}),p(\tilde{\bm{z}}|H_{1}))=N\times\left[\left(\frac{P_{D}\alpha_{F}}{P_{F}\alpha_{D}}+\frac{P_{F}\alpha_{D}}{P_{D}\alpha_{F}}\right)-2\right]$,
which represents a lower-bound for the system error probability \cite{Poor1994},
thus enforcing a zero Bayesian error exponent. Differently, the Neyman-Pearson
error exponent is given by $\lim_{K\rightarrow+\infty}-\frac{\ln\left[1-P_{D_{0}}(\bar{\gamma},K)\right]}{K}$,
under $P_{F_{0}}(\bar{\gamma},K)\leq\alpha$. If we choose $\bar{\gamma}{}_{\alpha}$
such that $P_{F_{0}}^{*}(\bar{\gamma}{}_{\alpha})=\alpha$, then $\lim_{K\rightarrow+\infty}-\ln\left[1-P_{D_{0}}(\bar{\gamma}{}_{\alpha},K)\right]=-\ln\left[1-P_{D_{0}}^{*}(\bar{\gamma}{}_{\alpha})\right]<+\infty$,
giving again a zero error exponent. 

Note that the performance in TPC scenarios differ through the ratio
$(\alpha_{F}/\alpha_{D})<1$ (cfr. Eqs.~(\ref{eq:asymptotic_perf_PF0})
and (\ref{eq:asymptotic_perf_PD0}) with Eqs. (\ref{eq:asymptotic_perf_PF0_TPC})
and (\ref{eq:asymptotic_perf_PD0_TPC})) which represents the \emph{performance
reduction factor }w.r.t. IPC scenarios. Note that $\alpha_{F}/\alpha_{D}$:
$(i)$ is an increasing function of the ratio $\frac{\sigma_{h}^{2}}{\sigma_{w}^{2}}$
(i.e. the received SNR), with limiting value equal to one; $(ii)$
is a decreasing function of $N$, meaning a diverging separation in
performance between IPC and TPC as $N$ increases.

\begin{figure}
\centering{}\includegraphics[scale=0.45]{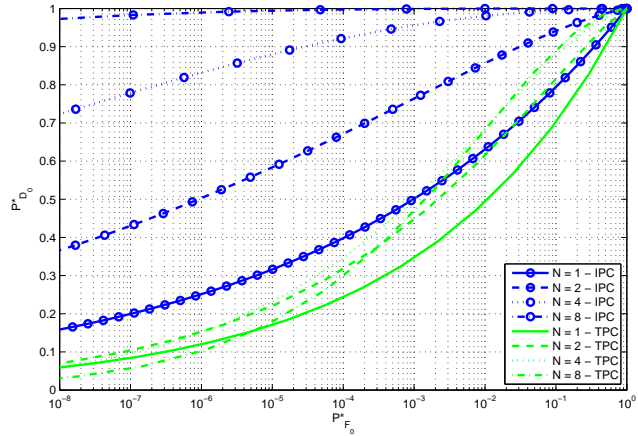}\caption{Effect of diversity $N$ on large-system ROC $(P_{D_{0}}^{*},P_{F_{0}}^{*})$
under both IPC and TPC; WSN with sensor characteristics $(P_{D,k},P_{F,k})=(P_{D},P_{F})=(0.5,0.05)$;
$(\sigma_{h}^{2}/\sigma_{w}^{2})_{dB}=15$. \label{fig:Diversity-effect-large system performances}}
\end{figure}

The diversity affects in a significant way the large-system probabilities
of detection and false alarm, under an IPC, by shifting the Receiver
Operating Characteristic (ROC) toward the upper-left corner, as shown
in Fig. \ref{fig:Diversity-effect-large system performances}, meaning
a performance improvement. Differently, it can be seen how a different
effect is present in the TPC case, where an increase of $N$ does
not always coincide with performance improvement, but rather an optimal
$N$, depending on $(P_{D},P_{F},\frac{\sigma_{h}^{2}}{\sigma_{w}^{2}})$,
exists. This effect was already noticed in \cite{Berger2009} and
it is due to non-coherent combining loss of branch contributions.

\begin{figure}
\centering{}\includegraphics[width=0.55\columnwidth]{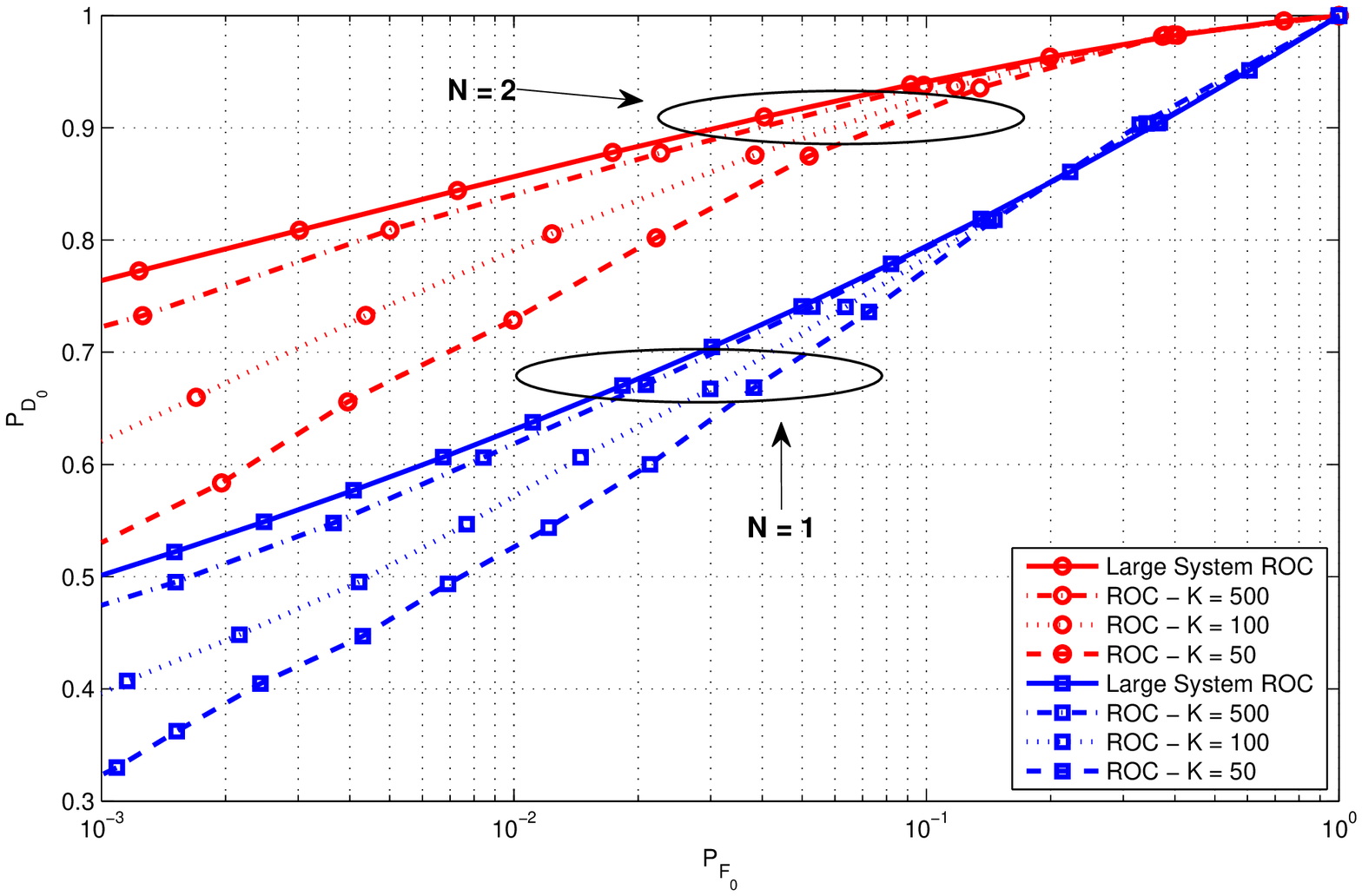}\caption{ROC comparison: large system vs finite number of sensors ($K\in\{50,100,500)$)
under IPC. $N\in\{1,2\}$, $(\sigma_{h}^{2}/\sigma_{w}^{2})_{dB}=15$.\label{fig:IPC_comparison}}
\end{figure}

\begin{figure}
\centering{}\includegraphics[width=0.55\columnwidth]{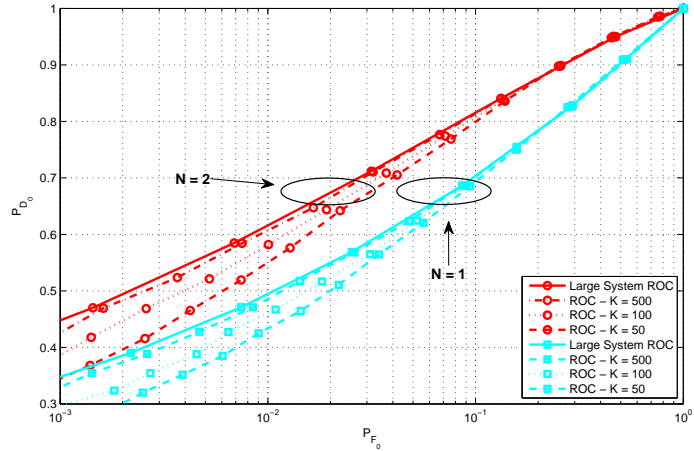}\caption{ROC comparison: large system vs finite number of sensors ($K\in\{50,100,500)$)
under TPC. $N\in\{1,2\}$, $(\sigma_{h}^{2}/\sigma_{w}^{2})_{dB}=15$.\label{fig:TPC_comparison}}
\end{figure}

Finally, in Figs. \ref{fig:IPC_comparison} and \ref{fig:TPC_comparison}
we verify, through simulations, the convergence of the ROC to the
large system expression ($K\rightarrow+\infty$) given by Eqs. (\ref{eq:asymptotic_perf_PF0})
and (\ref{eq:asymptotic_perf_PD0}) (resp. Eqs. (\ref{eq:asymptotic_perf_PF0_TPC})
and (\ref{eq:asymptotic_perf_PD0_TPC})), under IPC (resp. TPC). It
is apparent that the convergence under the TPC is faster w.r.t. the
IPC case, because in both cases the large system ROC expressions rely
on the Gaussian approximation of the Gaussian mixture given by Eq.
(\ref{eq:optimum_llr}). For such a reason, for a given $K$, imposing
a TPC on the WSN assures a better matching w.r.t. to the IPC case,
since all the components of the mixture will be more concentrated.

\section{Concluding Remarks\label{sec:Concluding-Remarks}}

In this correspondence we showed the optimality of the received-energy
test for decision fusion performed over a non-coherent diversity MAC
with conditionally i.n.i.d. sensor decisions. We derived a sufficient
condition on the LLR of the number of active sensors which can be
applied to test the received-energy optimality in WSN with conditionally
correlated sensor decisions. Finally, we showed, through analytical
results, how the diversity in a WSN with conditionally i.i.d sensor
decisions affects the large-system performance under both IPC and
TPC.

\section{Acknowledgement}

The authors would like to thank Prof. P. K. Willett at University
of Connecticut, US, for helpful discussions, as well as the associate
editor and the anonymous reviewers for their valuable comments that
allowed to improve significantly the correspondence.

\appendices{}

\section{Proof of Proposition 1 \label{sec:Appendix_Proposition-Suffcond}}

We prove in this section that Eq.~(\ref{eq:RFS-LLRmonotonicity})
is a sufficient condition for the optimality of $\psi$ test. From
Eq.~(\ref{eq:optimum_llr}), looking at the LLR as a function of
$\psi$, we get:

\begin{align}
\frac{\partial\Lambda_{opt}(\psi)}{\partial\psi} & =\frac{1}{\alpha(\psi)}\left[\sum_{\ell_{1}=0}^{K}\frac{\partial g(\psi,\ell_{1})}{\partial\psi}P(\ell_{1}|H_{1})\times\sum_{\ell_{2}=0}^{K}g(\psi,\ell_{2})P(\ell_{2}|H_{0})\right]\nonumber \\
 & -\frac{1}{\alpha(\psi)}\left[\sum_{\ell_{2}=0}^{K}\frac{\partial g(\psi,\ell_{2})}{\partial\psi}P(\ell_{2}|H_{0})\times\sum_{\ell_{1}=0}^{K}g(\psi,\ell_{1})P(\ell_{1}|H_{1})\right]\label{eq:Prop1_firstcondition}
\end{align}
 where we denoted $g(\psi,\ell)\triangleq\frac{1}{(\sigma_{w}^{2}+\ell\sigma_{h}^{2})^{N}}\exp\left(-\frac{\psi}{\sigma_{w}^{2}+\ell\sigma_{h}^{2}}\right)$
and $\alpha(\psi)$ indicates a positive function of $\psi$ (i.e.
$\alpha(\psi)>0$, $\forall\psi\in\mathbb{R}^{+}$) . Strictly increasing
property of LLR is guaranteed if $\frac{\partial\Lambda_{opt}(\psi)}{\partial\psi}>0$,
$\forall\psi\in\mathbb{R}^{+}$, thus manipulations from Eq.~(\ref{eq:Prop1_firstcondition})
lead to 
\begin{align}
\sum_{\ell_{1}=1}^{K}\sum_{\ell_{2}=0}^{\ell_{1}-1}k(\ell_{1},\ell_{2})\times\left[\frac{\partial g(\psi,\ell_{1})}{\partial\psi}g(\psi,\ell_{2})-\frac{\partial g(\psi,\ell_{2})}{\partial\psi}g(\psi,\ell_{1})\right] & >0\label{eq:first_ineq_monotonicity}
\end{align}
where $k(\ell_{1},\ell_{2})\triangleq\left[P(\ell_{1}|H_{1})P(\ell_{2}|H_{0})-P(\ell_{2}|H_{1})P(\ell_{1}|H_{0})\right]$.
In deriving Eq.~(\ref{eq:first_ineq_monotonicity}) we could express
the double sums in Eq.~(\ref{eq:Prop1_firstcondition}) as a function
only of the indices $\ell_{1}>\ell_{2}$, since the term in bracket
in Eq.~(\ref{eq:first_ineq_monotonicity}) equals to zero when $\ell_{1}=\ell_{2}$.
Noting that $\frac{\partial g(\psi,\ell)}{\partial\psi}=-\frac{1}{(\sigma_{w}^{2}+\ell\sigma_{h}^{2})}g(\psi,\ell)$
the condition is rewritten as
\begin{align}
\sum_{\ell_{1}=1}^{K}\sum_{\ell_{2}=0}^{\ell_{1}-1}k(\ell_{1},\ell_{2})g(\psi,\ell_{1})g(\psi,\ell_{2})\left[\frac{\sigma_{h}^{2}(\ell_{1}-\ell_{2})}{(\sigma_{w}^{2}+\ell_{1}\sigma_{h}^{2})(\sigma_{w}^{2}+\ell_{2}\sigma_{h}^{2})}\right] & >0\label{eq:final_llr_monotonicity_test}
\end{align}

Since both $g(\psi,\ell)$ and the term in square brackets are positive
(note that indices in the sums are such that $\ell_{1}>\ell_{2}$),
the term $k(\ell_{1},\ell_{2})$ is responsible for the sign of each
term in the sum. Then a \emph{sufficient condition} for Eq.~(\ref{eq:final_llr_monotonicity_test})
is obtained assuming that each of those terms is positive. This is
achieved if the following property holds
\begin{eqnarray}
k(\ell_{1},\ell_{2}) & > & 0,\quad\ell_{1}>\ell_{2}.\label{eq:monotonicity_LLR-l1l2}
\end{eqnarray}
 It is easy to demonstrate that the condition $k(\ell,\ell-1)>0$,
$\ell\in\mathcal{L}\backslash\{0\}$, representing the strictly increasing
property of $\ell$-LLR, i.e. $\frac{P(\ell|H_{1})}{P(\ell|H_{0})}>\frac{P(\ell-1|H_{1})}{P(\ell-1|H_{0})}$,
is equivalent to Eq.~(\ref{eq:monotonicity_LLR-l1l2}). In fact Eq.~(\ref{eq:monotonicity_LLR-l1l2})
implies that $\ell$-LLR is strictly increasing; this is verified
just substituting $\ell_{2}=\ell_{1}-1$. Differently, we can show
that $\ell$-LLR strictly increasing property implies Eq.~(\ref{eq:monotonicity_LLR-l1l2})
by constructing the chain of inequalities $\frac{P(\ell_{1}|H_{1})}{P(\ell_{1}|H_{0})}>\frac{P(\ell_{1}-1|H_{1})}{P(\ell_{1}-1|H_{0})}>\cdots>\frac{P(\ell_{2}|H_{1})}{P(\ell_{2}|H_{0})}$,
all deriving from $\ell$-LLR strictly increasing property.

\section{Proof of Theorem \ref{thm:RFS_monotonicity}\label{sec:Proposition_Monotonic LLR}}

We prove the strictly increasing property of $\ell$-LLR by induction.
Let us assume there exists a set of $(t-1)$ sensors with local performances
$(P_{D,k},P_{F,k})$, $k\in\{1,\ldots,t-1\}$. The number of active
sensors in this case is denoted $\ell_{t-1}\triangleq\sum_{k=1}^{t-1}x_{k}$,
$\ell_{t-1}\in\mathcal{L}_{t-1}$. We denote the probability of $\ell$
active sensors \emph{out-of-($t-1$),} given $H_{i}$, as $P_{t-1}(\ell|H_{i})$,
$H_{i}\in\mathcal{H}$ and the corresponding $\ell$-LLR as $\lambda_{t-1}(\ell)$.

\emph{Initialization}: the strictly increasing property of $\ell$-LLR
in single sensor case $\lambda_{1}(\ell_{1})>\lambda_{1}(\ell_{1}-1)$,
$\ell_{1}\in\mathcal{L}_{1}\backslash\{0\}$, is straightly verified
when $P_{D,1}>P_{F,1}$.

\emph{Induction: }Let us assume that for a specific configuration
of $(t-1)$ sensors the $\ell$-LLR $\lambda_{t-1}(\ell_{t-1})$ satisfies
the strictly increasing property, that is $\lambda_{t-1}(\ell_{t-1})>\lambda_{t-1}(\ell_{t-1}-1)$,
$\ell_{t-1}\in\mathcal{L}_{t-1}\backslash\{0\}$. If we add the $t$th
sensor satisfying $P_{D,t}>P_{F,t}$ and we prove that the new $\ell$-LLR
$\lambda_{t}(\ell_{t})>\lambda_{t}(\ell_{t}-1)$, $\ell_{t}\in\mathcal{L}_{t}\backslash\{0\}$,
i.e. it retains strictly increasing property, then the proof is complete. 

To proceed let us first define $a(\ell)\triangleq P_{t-1}(\ell|H_{1}$),
$b(\ell)\triangleq P_{t-1}(\ell|H_{0})$, $c(\ell)\triangleq P_{1}(\ell|H_{1})$
and $d(\ell)\triangleq P_{1}(\ell|H_{0})$. 

The number of sensors transmitting when the $t$th sensor is added
is then given by $\ell_{t}=\sum_{k=1}^{t}x_{k}=\ell_{t-1}+x_{t}$.
The pmfs $P_{t}(\ell_{t}|H_{0})$ and $P_{t}(\ell_{t}|H_{1})$ are
then given by \cite{Karr1993}
\begin{align}
P_{t}(\ell_{t}|H_{0})=(b*d)(\ell_{t})\qquad P_{t}(\ell_{t}|H_{1}) & =(a*c)(\ell_{t})
\end{align}
The LLR strictly increasing condition is then formulated as follows
\begin{equation}
\exp\left[\lambda_{t}(\ell_{t})\right]=\frac{(a*c)(\ell_{t})}{(b*d)(\ell_{t})}>\frac{(a*c)(\ell_{t}-1)}{(b*d)(\ell_{t}-1)}=\exp\left[\lambda_{t}(\ell_{t}-1)\right]\label{eq:LLR_monotonicity_condition}
\end{equation}
By exploiting the support set of $c(\ell)$ and $d(\ell)$ we can
rewrite Eq.~(\ref{eq:LLR_monotonicity_condition}) as follows
\begin{align}
\frac{\sum_{k\in\{0,1\}}c(k)a(\ell_{t}-k)}{\sum_{k\in\{0,1\}}d(k)b(\ell_{t}-k)} & >\frac{\sum_{k\in\{0,1\}}c(k)a(\ell_{t}-1-k)}{\sum_{k\in\{0,1\}}d(k)b(\ell_{t}-1-k)}
\end{align}
where obviously $a(t)=b(t)=0$. Exploiting $c(0)+c(1)=(1-P_{D,t})+P_{D,t}=1$,
$d(0)+d(1)=(1-P_{F,t})+P_{F,t}=1$, we obtain
\begin{align}
\frac{[1-c(1)]a(\ell_{t})+c(1)a(\ell_{t}-1)}{[1-d(1)]b(\ell_{t})+d(1)b(\ell_{t}-1)} & >\frac{[1-c(1)]a(\ell_{t}-1)+c(1)a(\ell_{t}-2)}{[1-d(1)]b(\ell_{t}-1)+d(1)b(\ell_{t}-2)}\label{eq:Difference of LLR}
\end{align}
The condition expressed in Eq.~(\ref{eq:Difference of LLR}) can
be rewritten as:
\begin{multline}
\left\{ [1-c(1)][1-d(1)][a(\ell_{t})b(\ell_{t}-1)-a(\ell_{t}-1)b(\ell_{t})]\right\} +\left\{ c(1)d(1)[a(\ell_{t}-1)b(\ell_{t}-2)-a(\ell_{t}-2)b(\ell_{t}-1)]\right\} +\\
+\left\{ c(1)[1-d(1)][a(\ell_{t}-1)b(\ell_{t}-1)-a(\ell_{t}-2)b(\ell_{t})]\right\} -\left\{ [1-c(1)]d(1)[a(\ell_{t}-1)b(\ell_{t}-1)-a(\ell_{t})b(\ell_{t}-2)]\right\} >0\label{eq:Total condition}
\end{multline}
Since $\frac{a(\ell_{t})}{b(\ell_{t})}>\frac{a(\ell_{t}-1)}{b(\ell_{t}-1)}>\frac{a(\ell_{t}-2)}{b(\ell_{t}-2)}$,
we have that: 
\begin{align}
\left[a(\ell_{t})b(\ell_{t}-1)-a(\ell_{t}-1)b(\ell_{t})\right] & >0\qquad\left[a(\ell_{t}-1)b(\ell_{t}-2)-a(\ell_{t}-2)b(\ell_{t}-1)\right]>0\label{eq:inequalities_LLR1}
\end{align}
\begin{equation}
\left[a(\ell_{t})b(\ell_{t}-2)-a(\ell_{t}-2)b(\ell_{t})\right]>0\label{eq:inequalities_LLR_2}
\end{equation}
The condition in Eq.~(\ref{eq:Total condition}) is satisfied since
positivity of the first two terms follows from the inequalities in
Eq.~(\ref{eq:inequalities_LLR1}), and the difference of the third
and fourth terms in Eq.~(\ref{eq:Total condition}) is positive because
$c(1)[1-d(1)]>[1-c(1)]d(1)$ (since $P_{D,t}>P_{F,t}$) and exploiting
the inequality in Eq.~(\ref{eq:inequalities_LLR_2}). This concludes
the proof.

\section{Proof of Theorem \ref{thm:Theorem4}\label{sec:Proof-of-Theorem 4}}

The proof follows in the first part similar steps as in \cite{Li2011};
for this reason we will only sketch it and underline the substantial
differences. We use here the \emph{characteristic function }of the
vector $\bm{z}|H_{i}$, $i\in\{1,2\}$, denoted as $\Phi_{\bm{z}|H_{i}}(\bm{t})$,
to easily evaluate the limit for $K\rightarrow+\infty$. We then use
this result, in conjunction with \emph{Levy's Continuity Theorem}
\cite{Karr1993} to demonstrate the convergence in distribution of
large-system $p(\bm{z}|H_{i})$. Let us now write the characteristic
function of $\bm{z}|H_{0}$ as a function of $\bm{t=}\left(\bm{t}_{1}^{t},\bm{t}_{2}^{t}\right)^{t}$:
\begin{align}
\Phi_{\bm{z}|H_{0}}(\bm{t}) & =\mathbb{E}_{\bm{z}|H_{0}}\{\exp(j\bm{t}_{1}^{t}\bm{z}_{1}+j\bm{t}_{2}^{t}\bm{z}_{2})\}=\dotsint\exp(j\bm{t}_{1}^{t}\bm{z}_{1}+j\bm{t}_{2}^{t}\bm{z}_{2})\times\sum_{\ell=0}^{K}p(\bm{z}_{1},\bm{z}_{2}|\ell)P(\ell|H_{0})d\bm{z}_{1}d\bm{z}_{2}
\end{align}
where $\bm{z}_{1}\triangleq\Re\{\bm{z}\}$ and $\bm{z}_{2}\triangleq\Im\{\bm{z}\}$
(with $\bm{t}_{i}$, $i\in\{1,2\}$, representing the index-corresponding
dual vectors over Fourier domain). Following analogous steps as in
\cite{Li2011}, exploiting: $i)$ conditional independence assumptions
such as $p(\bm{z}_{1},\bm{z}_{2}|\ell)=p(\bm{z}_{1}|\ell)p(\bm{z}_{2}|\ell)$
and $p(\bm{z}_{i}|\ell)=\prod_{s=1}^{N}p(z_{i,s}|\ell)$, $i\in\{1,2\}$;
$ii)$ the characteristic function of $x\sim\mathcal{N}(0,\sigma^{2})$
is given by $\Phi_{x}(t)=\exp(-\frac{\sigma^{2}t^{2}}{2})$ \cite{Karr1993};
we get
\begin{align}
\Phi_{\bm{z}|H_{0}}(\bm{t}) & =\sum_{\ell=0}^{K}P(\ell|H_{0})\times\exp\left[-\frac{1}{4}\sum_{s=1}^{N}(t_{1,s}^{2}+t_{2,s}^{2})\left(\frac{\ell}{NKP_{F}}+\frac{\sigma_{w}^{2}}{\sigma_{h}^{2}KP_{F}}\right)\right]\nonumber \\
 & =\exp\left[-\frac{1}{4}\sum_{s=1}^{N}(t_{1,s}^{2}+t_{2,s}^{2})\frac{\sigma_{w}^{2}}{\sigma_{h}^{2}KP_{F}}\right]\times\left\{ P_{F}\exp\left[-\frac{1}{4}\sum_{s=1}^{N}\frac{(t_{1,s}^{2}+t_{2,s}^{2})}{NKP_{F}}\right]+(1-P_{F})\right\} ^{K}
\end{align}

where in the last line we exploited $\ell|H_{0}\sim\mathcal{B}(K,P_{F})$.
Also, exploiting similar noteworthy limits as in \cite{Li2011}, eventually
we have that $\lim_{K\rightarrow+\infty}\Phi_{\bm{z}|H_{0}}(\bm{t})=\exp\left[-\frac{1}{2}\sum_{s=1}^{N}\frac{(t_{1,s}^{2}+t_{2,s}^{2})}{2N}\right]$.
Applying the Continuity Theorem \cite{Karr1993}, we obtain $\bm{z}|H_{0}\overset{d}{\rightarrow}\mathcal{N}_{\mathbb{C}}\left(\bm{0}_{N},\frac{1}{N}\bm{I}_{N}\right)$.
In a similar way it can be shown that $\bm{z}|H_{1}\overset{d}{\rightarrow}\mathcal{N}_{\mathbb{C}}\left(\bm{0}_{N},\frac{P_{D}}{P_{F}N}\bm{I}_{N}\right)$. 

The last part consists in proving Eqs.~(\ref{eq:asymptotic_perf_PF0})
and (\ref{eq:asymptotic_perf_PD0}). The large-system probabilities
of false alarm and detection can be expressed in the equivalent form:
\begin{eqnarray}
P_{F_{0}}^{*}(\bar{\gamma}) & = & P(\left\Vert \bm{z}\right\Vert ^{2}\geq\bar{\gamma}|H_{0})=P\left(\frac{1}{2N}\xi\geq\bar{\gamma}|H_{0}\right)\\
P_{D_{0}}^{*}(\bar{\gamma}) & = & P(\left\Vert \bm{z}\right\Vert ^{2}\geq\bar{\gamma}|H_{1})=P\left(\frac{P_{D}}{2P_{F}N}\xi\geq\bar{\gamma}|H_{1}\right)
\end{eqnarray}
where $\xi\sim\chi_{(2N)}^{2}$. The probabilities are then easily
calculated evaluating the cumulative distribution function of $\xi$
\cite{Karr1993}:
\begin{equation}
P_{F_{0}}^{*}(\bar{\gamma})=\int_{2\bar{\gamma}N}^{+\infty}p(\xi)d\xi\qquad P_{D_{0}}^{*}(\bar{\gamma})=\int_{2\bar{\gamma}N\frac{P_{F}}{P_{D}}}^{+\infty}p(\xi)d\xi
\end{equation}
 which provides the result.

\bibliographystyle{IEEEtran}
\bibliography{IEEEabrv,sensor_networks}

\end{document}